\documentclass[sigconf]{acmart}

\usepackage[caption=false, font=footnotesize]{subfig}
\usepackage{paralist}
\usepackage[figure,table]{hypcap}
\usepackage{amsmath}
\usepackage{float}
\usepackage{algorithm}
\usepackage{algpseudocode}
\usepackage{array}

\usepackage[whole]{bxcjkjatype}

\usepackage{alphabeta}
\usepackage{arabtex}
\usepackage[LFE,LAE,LGR,T2A,T1]{fontenc}
\usepackage[greek, russian, main=english]{babel}

\AtBeginDocument{%
  \providecommand\BibTeX{{%
    \normalfont B\kern-0.5em{\scshape i\kern-0.25em b}\kern-0.8em\TeX}}}

\setcopyright{acmcopyright}
\copyrightyear{2024}
\acmYear{2024}
\acmDOI{10.1145/1122445.1122456}

\copyrightyear{2024}
\acmYear{2024}
\setcopyright{rightsretained}
\acmConference[GECCO '24 Companion]{Genetic and Evolutionary Computation Conference}{July 14--18, 2024}{Melbourne, VIC, Australia}
\acmBooktitle{Genetic and Evolutionary Computation Conference (GECCO '24 Companion), July 14--18, 2024, Melbourne, VIC, Australia}
\acmDOI{10.1145/3638530.3654432}
\acmISBN{979-8-4007-0495-6/24/07}

\begin{document}

\title{LVNS-RAVE: Diversified audio generation with RAVE and Latent Vector Novelty Search}

\author{
    Jinyue Guo
    }
\affiliation{%
    \department{RITMO, Department of Musicology}
  \institution{University of Oslo}
  \city{Oslo}
  \country{Norway}
}
\email{jinyue.guo@imv.uio.no}
\orcid{0009-0003-7438-1483}

\author{
    Anna-Maria Christodoulou
    }
\affiliation{%
    \department{RITMO, Department of Musicology}
  \institution{University of Oslo}
  \city{Oslo}
  \country{Norway}
}
\email{a.m.christodoulou@imv.uio.no}
\orcid{0009-0003-7063-2062}

\author{
    Bálint Laczkó
    }
\affiliation{%
    \department{RITMO, Department of Musicology}
  \institution{University of Oslo}
  \city{Oslo}
  \country{Norway}
}
\email{balint.laczko@imv.uio.no}
\orcid{0009-0001-8337-2509}

\author{
    Kyrre Glette
    }
\affiliation{%
    \department{RITMO, Department of Informatics}
  \institution{University of Oslo}
  \city{Oslo}
  \country{Norway}
}
\email{kyrrehg@ifi.uio.no}
\orcid{0000-0003-3550-3225}

\renewcommand{\shortauthors}{Guo, et al.}

\begin{abstract}
    Evolutionary Algorithms and Generative Deep Learning have been two of the most powerful tools for sound generation tasks. However, they have limitations: Evolutionary Algorithms require complicated designs, posing challenges in control and achieving realistic sound generation. Generative Deep Learning models often copy from the dataset and lack creativity. In this paper, we propose \textit{LVNS-RAVE}, a method to combine Evolutionary Algorithms and Generative Deep Learning to produce realistic and novel sounds. We use the RAVE model as the sound generator and the VGGish model as a novelty evaluator in the Latent Vector Novelty Search (LVNS) algorithm. The reported experiments show that the method can successfully generate diversified, novel audio samples under different mutation setups using different pre-trained RAVE models. The characteristics of the generation process can be easily controlled with the mutation parameters. The proposed algorithm can be a creative tool for sound artists and musicians.
\end{abstract}

\ccsdesc[500]{Applied computing~Sound and music computing}
\ccsdesc[300]{Information systems~Multimedia content creation}

\keywords{Neural audio synthesis, variational autoencoder, latent vector evolution}

\begin{teaserfigure}
 \centerline{\includegraphics[width=0.86\linewidth]{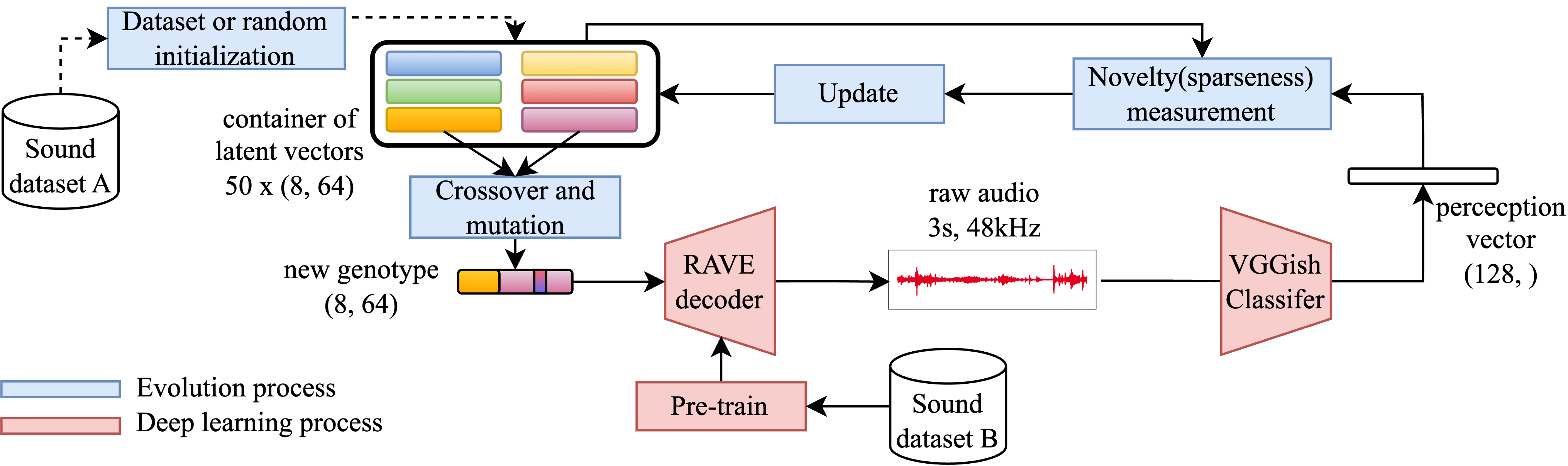}}
 \caption{The overall workflow of LVNS-RAVE: the red blocks show the Deep Learning process for high-quality audio generation and perception estimation, and the blue blocks show the Novelty Search process that looks for diversified, novel samples.}
 \label{fig:teaser}
\end{teaserfigure}


\maketitle

\section{Introduction}\label{sec:introduction}

Since the introduction of \textit{wavenet}~\cite{oord_wavenet_2016}, generative audio Deep Learning models~\cite{garcia_vampnet_2023, schneider_mousai_2023} have been developing rapidly. By efficiently modeling the temporal distribution of audio samples, these models can generate realistic, high-fidelity audio outputs. The RAVE model~\cite{caillon_rave_2021}, with improvements of the Variational Autoencoder (VAE), excels in balancing fidelity and computational complexity. However, the nature of these probability models limits their ability to generate samples similar to their training dataset, often lacking diversity and novelty.

Evolutionary Algorithms provide an alternative for audio generation. They can generate novel audio samples without the limitation of pre-defined datasets. 
Parameter values for synthesizer models are evolved in~\cite{dahlstedt_evolution_2007} and~\cite{mcdermott_evolutionary_2008}, and compositional pattern-producing networks (CPPNs) are used for synthesis with interactive evolution in~\cite{jonsson_interactively_2015}. 
The Sound Innovation Engines approach~\cite{jonsson-sie} employs Quality-Diversity search and automatic fitness evaluation. However, in all of these approaches the quantitative quality measurement of samples remains a challenge. 

To combine the fidelity of Deep Learning models and the diversity of Evolutionary Algorithms, Bontrager et al.~\cite{bontrager_deepmasterprints_2018} introduced Latent Variable Evolution (LVE). Using the VAE model, they generated \textit{MasterPrints}, a set of natural and synthetic fingerprints. The VAE learned a latent representation of the original fingerprint images, ensuring any latent vector could be decoded into a meaningful fingerprint image. The LM-MA-ES algorithm then evolved latent vectors to match most fingerprint images. LVE was also applied to other fields such as game level generation~\cite{sarkar_generating_2021, tanabe_level_2021}, illustrating LVE's potential for generating diversified and high-quality samples.

In this paper, we make an initial exploration of applying LVE to audio generation. Using Novelty Search to evolve the latent vectors of the RAVE model and a deep learning-based audio distance measure, we propose the LVNS-RAVE~\footnote{Code and audio samples can be found at \url{https://github.com/fisheggg/LVNS-RAVE}.} method that can generate realistic and diversified audio samples.

\section{Methods}
    

The RAVE model has an autoencoder structure with two parts of networks: the encoder and the decoder. The encoder can be seen as a compression network that extracts meaningful information from input audio waveforms to form the latent vectors $\mathbf{s}_i \in \mathbb{R}^{d \times l}$, where $d$ represents the RAVE latent dimension, usually between 8 and 20 according to the pre-trained model, and $l$ represents the sequence length that is fixed at 64. In our experiment, we utilize the latent vectors as genotypes and use the RAVE decoder network to map them back to audio waveforms or phenotypes. The variational modeling and generative adversarial fine-tuning of RAVE create a smooth latent space with semantic meanings so that all vectors in the latent space are mapped to meaningful phenotypes and that vectors with smaller distances are more similar. These properties make it ideal for the evolutionary algorithm to explore the latent space and find diversified phenotypes.

We chose the Novelty Search algorithm~\cite{lehman_abandoning_2011, lehman_evolving_2011} to evolve the latent vectors of RAVE. This decision stems from our primary objective to generate diversified and realistic audio samples. Common Evolutionary Algorithms (CMA-ES or LM-MA-ES) geared towards optimizing a single metric, do not align with our diversity-centric goal. On the other hand, although proven powerful in generating realistic audio samples from a given training set, the RAVE model often lacks diversity and novelty in their output. Since defining a qualitative measurement for sound is difficult, we want to use the realistic RAVE models to guarantee quality. At the same time, the evolutionary part can focus on improving diversity. We modified the Novelty Search algorithm's original crossover and selection process to adapt for the RAVE latent vector sequences. The LVNS-RAVE process, depicted in Figure \ref{fig:teaser}, includes the following four stages:
    
\paragraph{Initialization}

We define the container as a set $\mathbf{C} = \{\mathbf{s}_1, \mathbf{s}_2, \dots \mathbf{s}_n\}$, with $n$ denoting the sample count. Container initialization involves two methods: drawing samples from a multivariate normal distribution or extracting latent vectors from a dataset. The latter involves randomly selecting samples and using the RAVE encoder to convert audio waveforms into latent embedding sequences.

\paragraph{Crossover and Mutation}

We define genotypes as matrices in our experiments; $\mathbf{s}_i\in \mathbb{R}^{d\times l}$.  Using columns as the basic elements for crossover and mutation processes, each column represents a latent vector of an audio frame.
%
To generate a new genotype, we randomly select two samples $\mathbf{s}_i$ and $\mathbf{s}_j$ from the container $\mathbf{C}$ as parents, with all samples having equal possibilities. Then, we pick a random breakpoint $k$ and assign the columns before or at $m$ from $\mathbf{s}_i$, and columns after $m$ from $\mathbf{s}_j$, to the child $\mathbf{s}^*$. After the crossover, we select one column in $\mathbf{s}^*$ to add noise from a normal distribution as the mutation. We keep generating new breeds and add them to the set $\mathbf{C}_{new}$, until the set has $N$ elements.
    
    
\paragraph{Novelty Evaluation}

To measure the novelty of newly generated breeds, we use the VGGish model\cite{hershey_cnn_2017} that was pre-trained on AudioSet\cite{gemmeke_audio_2017} as a general auditory measurement model. The 128-dimension vector before the classification head serves as the \textit{perception vector}. Since the VGGish model was trained on general audio samples, unlike the specifically trained RAVE models, we expect it to output the ``perceptual understanding" of the evolved samples. We use the Euclidean distance between the VGGish embeddings of the two audio samples as their similarity score, with lower distances meaning more similar samples. The distance between genotype $\mathbf{s_i}$ and $\mathbf{s_j}$ is given by:

\begin{align}
    \mathrm{dist}(\mathbf{s}_i, \mathbf{s}_j) = ||\mathrm{VGG}\mathrm{ish}(\mathrm{RDec}(\mathbf{s}_i)) - \mathrm{VGGish}(\mathrm{RDec}(\mathbf{s}_j))||
\end{align}

where $RDec()$ represents the pre-trained RAVE decoder.
    
Then, we measure the novelty by calculating the sparseness $\rho$ of each sample, which is the average distance of each sample to its k-nearest neighbors with the container $\mathbf{C}$ and the set of newly generated breeds $\mathbf{C}_{new}$:

\begin{align}
\begin{split}
    \rho(\mathbf{s}_i) &= \frac1k \sum_{j\in \mathbf{U}}\mathrm{dist}(\mathbf{s}_i, \mathbf{s}_j),\\
    \mathrm{where\ } \mathbf{U} &= \mathrm{knn}(\mathbf{s}_i, \mathbf{C} \bigcup \mathbf{C}_{new})
\end{split}
\end{align}
    
$\mathrm{knn}(\mathrm{s}, \mathrm{C})$ represents the k-nearest neighbors of point $\mathbf{s}$ in the data collection set $C$. We use $\mathbf{k=50}$ in all following experiments.
    
\paragraph{Selection and Update}

After calculating the sparseness of every sample in $\mathbf{C}_{new}$, we define the new container $\mathbf{C}^*$ with samples that has the highest sparseness values:

\begin{align}
    \mathbf{C}^* = \mathrm{argmax}_{\mathbf{C}'\subset \mathbf{C}\bigcup\mathbf{C_{new}}, |\mathbf{C}'|=N}\sum_{\mathbf{s}_i\in \mathbf{C}'}\rho(\mathbf{s}_i)
\end{align}

where $N$ is the size of the container.

\section{Experiments and evaluation}

We conducted the evolutionary experiments on three different pre-trained RAVE models\footnote{\url{https://acids-ircam.github.io/rave\_models\_download}}: \textit{vintage}, \textit{darbouka\_onnx}, and \textit{VCTK}. The \textit{vintage} model was trained on 80 hours of vintage music, the \textit{darbouka\_onnx} model was trained on 8 hours of darbouka recordings, and the \textit{VCTK} model was trained on CSTR's VCTK Corpus. We used the pytorch implementation\footnote{\url{https://github.com/harritaylor/torchvggish/tree/master}} of the VGGish model.
The evolution process was computed using a server with 8 CPUs and no GPU. Table \ref{tab:evo_ini_experiment_setup} shows the setups used in the experiments. 


\begin{table}[hb]
    \caption{Experiment setups of Latent Vector Novelty Search.}
    \label{tab:evo_ini_experiment_setup}
    \begin{tabular}{>{\centering\arraybackslash}p{0.1\linewidth}>{\centering\arraybackslash}p{0.15\linewidth}>{\centering\arraybackslash}p{0.15\linewidth}>{\centering\arraybackslash}p{0.1\linewidth}>{\centering\arraybackslash}p{0.25\linewidth}}
        \toprule
        \textbf{Setup No.} & \textbf{Number of Generations}& \textbf{Container Size}& \textbf{New Breed Size}& \textbf{Container Initialization Method}\\
        \midrule
        1 & 100 & 50 & 30 & random \\
        2 & 100 & 50 & 30 & dataset \\
        3 & 300 & 50 & 30 & dataset \\
        4 & 100 & 500 & 50 & dataset \\
        \bottomrule
    \end{tabular}
\end{table}

\begin{figure}[htbp]
    \centering
    \includegraphics{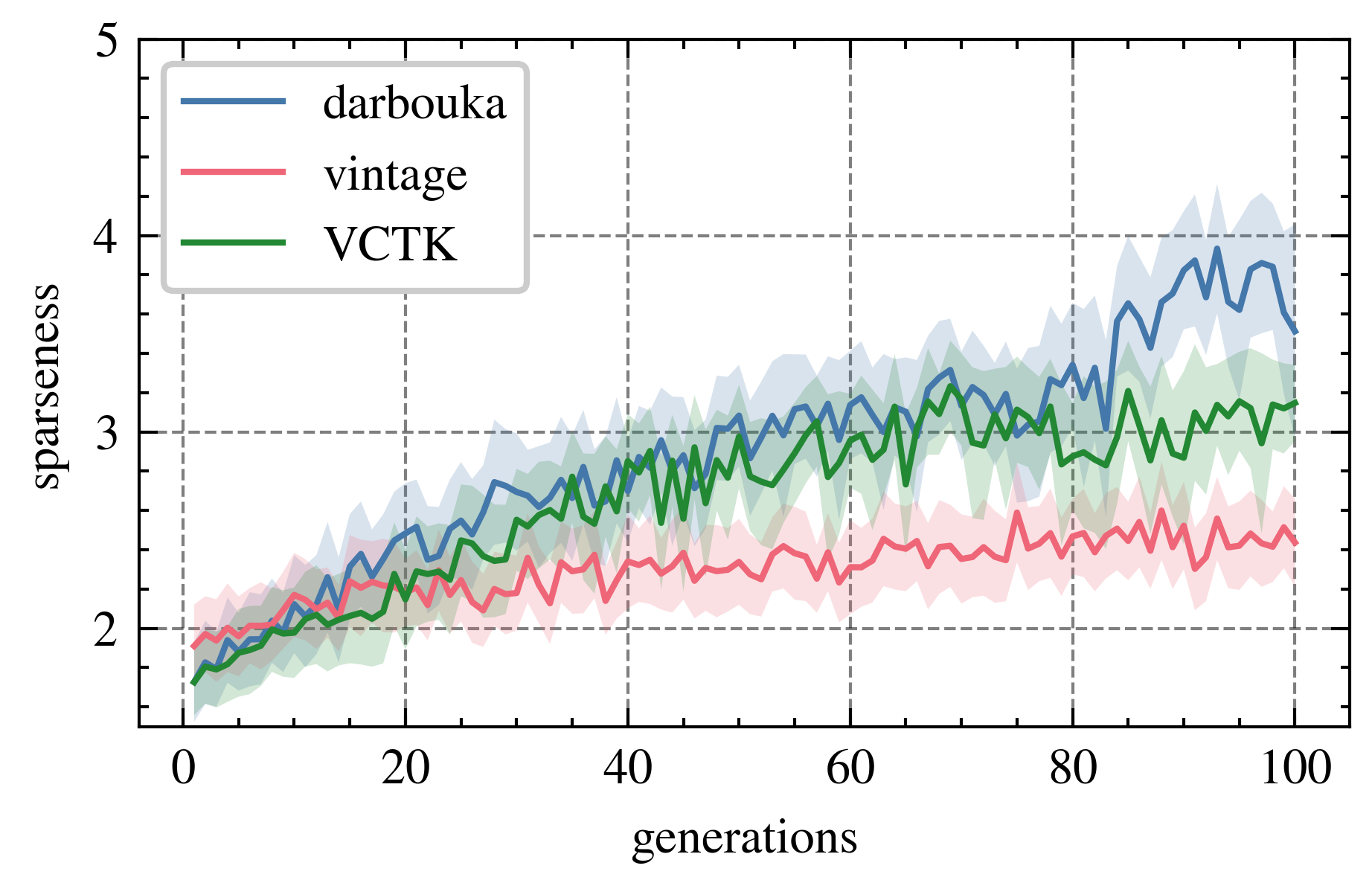}
    \caption{Experiment group 1: Sparseness of the containers during the evolution process of three different pre-trained models, all using experiment Setup 1. Lines show the mean sparseness; shadow areas show the standard deviation.}
    \Description{A line chart where the vertical axis is labeled "sparseness" and the horizontal axis is labeled "generations". There are three lines of different colors denoting mean values. These are overlaid on top of matching lighter-colored areas that denote standard deviation. The lines all originate roughly at the same point in the lower left corner but increasingly deviate as they go to the right.}
    \label{fig:sparseness_1}
\end{figure}

\begin{figure*}[htbp]
    \centering
    \includegraphics[width=0.8\linewidth]{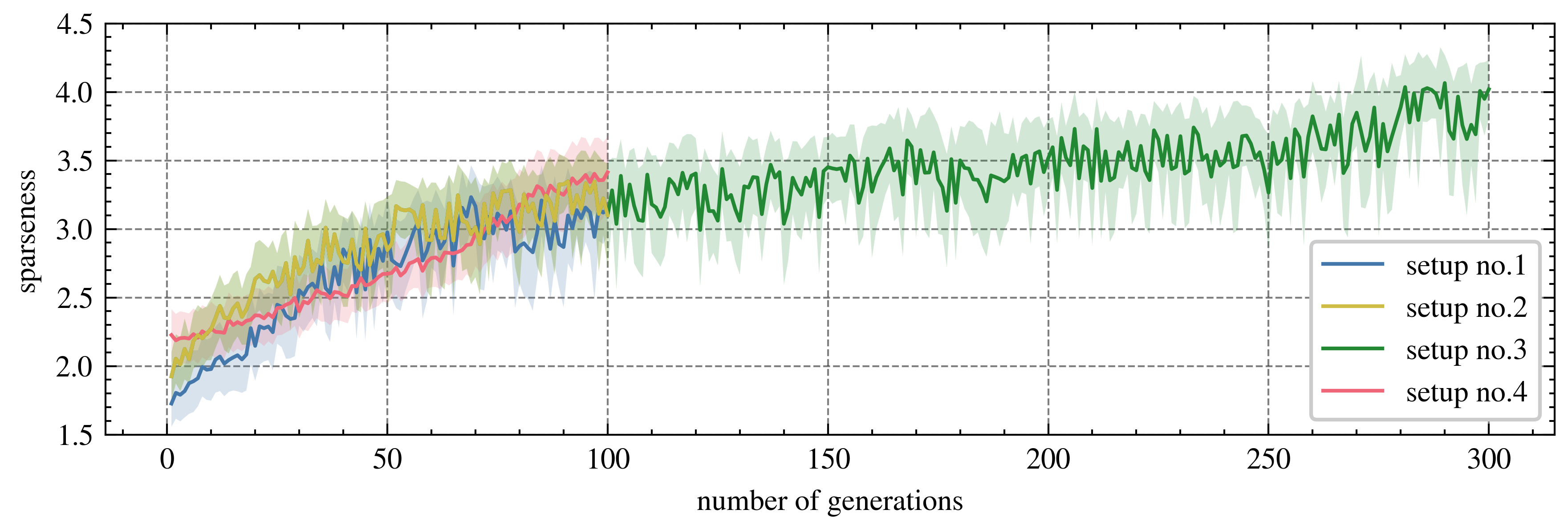}
    \caption{Experiment group 2: Sparseness of the containers during the evolution process using VCTK pre-train model, under 4 different experiment setups. Lines show the mean sparseness and shadow areas show the standard deviation.}
    \Description{A line chart where the vertical axis is labeled "sparseness" and the horizontal axis is labeled "generations". There are four lines of different colors denoting mean values. These are overlaid on top of matching lighter-colored areas that denote standard deviation. The lines all originate at roughly the same point in the lower left corner. Three lines only reach one-third of the horizontal span, and only the fourth line spans the available space. The lines stay fairly closely together. The longer line reaches the top-right corner in a round trace with a steeper slope on the left side and gradually flattens towards the right.}
    \label{fig:sparseness_2}
\end{figure*}

\begin{figure*}[htbp]
    \centering
    \includegraphics[width=0.75\linewidth]{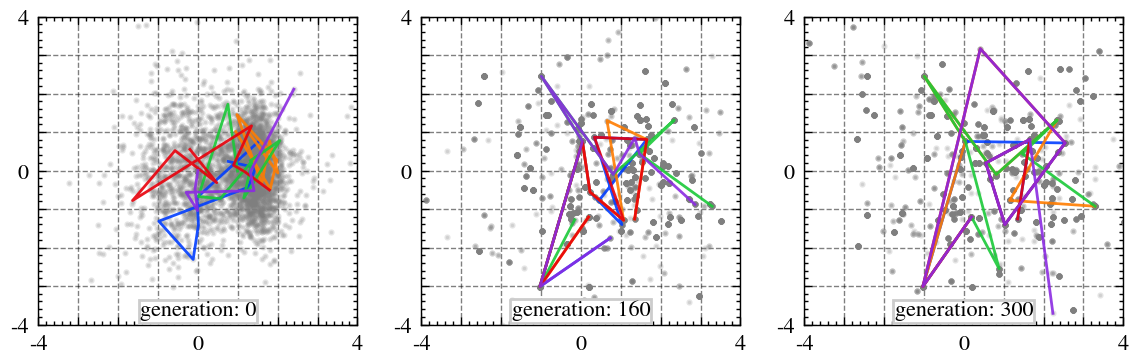}
    \caption{Plot of the first two dimensions of the container embedding sequences, generated using Setup 3. Colored lines show the trajectories of five sequences. Gray dots show the distribution of all other embedding vectors in the container.}
    \Description{Three square scatter plots in a row, where all axes are unlabeled and range from -4 to 4. The background of the scatter plots shows different distributions of gray point clouds, while in the foreground, there are 5 differently colored lines connecting some of the points. The plots are labeled "generation 0", "generation 160" and "generation 300", respectively. The lines in the left-most plot are more in the center, while in the right-most plot, they spread out more.}
    \label{fig:embeddings}
\end{figure*}

We conducted two groups of experiments: Group 1 used different pre-trained RAVE models under the same setup; Group 2 used the same \textit{VCTK} RAVE model under different setups. Figure \ref{fig:sparseness_1} and \ref{fig:sparseness_2} show the sparseness of the containers of each group along the evolutionary generations. Figure \ref{fig:embeddings} plots the distribution of embedding vectors during the evolution process in group 2, Setup 3, and highlights some of the trajectories. Due to time limitations, the results are from single runs to demonstrate insights, but more repetitions will be needed to ascertain statistically significant differences.

Figure \ref{fig:sparseness_1} shows the sparseness plot using different pre-trained RAVE models. Since the containers were randomly initialized, all experiments had an initial sparseness of around $1.8$. Over generations, the mean sparseness of the containers grew, while their standard deviations staid in the same range. This indicates that the VGGish evaluation model can create environmental pressure, leading to more diversified species. The growth rate varies but is present in different models. Lower growth rates do not necessarily correspond to less diversified samples since the space of the evaluation vectors is different for each model. For example, the \textit{vintage} model creates audio samples with similar frequency ranges, which the VGGish model perceived as less sparse. That does not imply that the evolution results of the \textit{vintage} experiment are less diversified than the other experiments. It is only meaningful to compare the relative levels of diversity between the initial and the final container of the same experiment.

Figure \ref{fig:sparseness_2} shows the sparseness of the four different setups using the same \textit{VCTK} model. In Setup 3, the growth of sparseness slowed down after the first 100 generations, but we did not observe a limit to the growth. Setup 4 used a larger container size of 500 instead of 50, with the number of new breeds being increased accordingly. The initial sparseness of Setup 4 is higher than the other setups, but there is no apparent difference after 100 generations. We observe a more linear growth of Setup 3 than the others. Compared to the random initialization of Setup 1, the other setups that use samples from the original dataset have larger initial sparseness. However, after 100 generations, there is no significant difference between the two initialization methods.

Figure \ref{fig:embeddings} plots the distribution of embedding vectors during evolution and highlights some trajectories. Over generations, the range of the embedding vectors expanded beyond the original dataset's distribution, while the directions and shapes of the embedding sequences have also become more diversified. We also observed much sparser embedding distributions since the first 40 generations.

\section{Conclusion and future work}

This paper introduced our initial effort to apply latent variable evolution to deep audio generation using Novelty Search on RAVE latent sequences. We used an external evaluation network to measure the similarity between samples and calculated the sparsenesses of each phenotype. Our exploratory experiments indicate that Latent Variable Evolution selecting for Novelty using a VGGish distance measure can lead to a notable increase in the diversity of the generated results. We tested the method over three pre-trained RAVE models and four experiment setups. The algorithm is flexible to produce high-quality sound samples with various characteristics, which is preferable for artistic purposes.

In future work, we will collect more data from multiple evolutionary runs and explore a broader range of hyperparameters.
Larger containers with thousands of samples could lead to different behaviors. Since the RAVE models use a sequence of latent vectors as the embeddings, we have adjusted the original crossover and mutation process. However, there might be better mutation methods that can further increase the diversity of phenotypes. Finally, quality evaluation has been a consistent problem for audio generation. Human assessment could be involved to provide better guidance to the generator.

\begin{acks}
This project is supported by the Research Council of Norway through projects 262762 (RITMO) and 324003 (AMBIENT). 
\end{acks}

\bibliographystyle{ACM-Reference-Format}
\bibliography{references,jonsson-sie}


\end{document}